\documentclass[runningheads]{llncs}

\usepackage{threeparttable}
\usepackage{adjustbox}
\usepackage{array}
\usepackage{multirow}
\usepackage{hyperref}
\usepackage{amsmath}
\usepackage{mathtools}
\usepackage{booktabs}
\usepackage{enumitem}
\usepackage[table,xcdraw]{xcolor}
\usepackage{url}
\usepackage{makecell}
\usepackage{threeparttable}
\usepackage{comment}

\usepackage{pifont}
\newcommand{\cmark}{\ding{51}}%
\newcommand{\xmark}{\ding{55}}%

\begin{document}

\title{EmLog: Tamper-Resistant System Logging for Constrained Devices with TEEs}
\titlerunning{EmLog: Tamper-Resistant Logging for Constrained Devices with TEEs}

\author{Carlton Shepherd \and Raja Naeem Akram \and Konstantinos Markantonakis}
\authorrunning{Shepherd, Akram, and Markantonakis} 
\institute{Smart Card and Internet of Things Security Centre, Information Security Group, Royal Holloway, University of London, Surrey, United Kingdom.\\
\email{\{carlton.shepherd.2014, r.n.akram, k.markantonakis\}@rhul.ac.uk}}
\tocauthor{Carlton Shepherd, Raja N. Akram and Konstantinos Markantonakis}

\maketitle  

\begin{abstract}
Remote mobile and embedded devices are used to deliver increasingly impactful services, such as medical rehabilitation and assistive technologies.  Secure system logging is beneficial in these scenarios to aid audit and forensic investigations particularly if devices bring harm to end-users.   Logs should be tamper-resistant in storage, during execution, and when retrieved by a trusted remote verifier.  In recent years, Trusted Execution Environments (TEEs) have emerged as the go-to root of trust on constrained devices for isolated execution of sensitive applications.   Existing TEE-based logging systems, however, focus largely on protecting server-side logs and offer little protection to constrained source devices.  In this paper, we introduce EmLog -- a tamper-resistant logging system for constrained devices using the GlobalPlatform TEE.  EmLog provides protection against complex software adversaries and offers several additional security properties over past schemes.  The system is evaluated across three log datasets using an off-the-shelf ARM development board running an open-source, GlobalPlatform-compliant TEE.  On average, EmLog runs with low run-time memory overhead (1MB heap and stack), 430--625 logs/second throughput, and five-times persistent storage overhead versus unprotected logs. 
\end{abstract}

\section{Introduction}
\label{sec:intro}

System logs record features such as user activity, resource consumption, peripheral use and error details.  Logs are also used to enforce user accountability and to establish audit trails for forensics, event reconstruction and intrusion detection~\cite{nist:logs}.  Consequently, logs are routinely targeted by attackers to conceal evidence of wrongdoing, and should be stored securely to preserve the auditability of a compromised system -- as recommended by NIST~\cite{nist:logs} and ISO 27001:2013~\cite{iso}.  Not only should logs be stored in a way that cryptographically preserves their confidentiality and integrity, but trusted computing primitives, e.g. Trusted Platform Modules (TPMs), have been identified as desirable in existing proposals~\cite{bock,sinha:tpm}.  Such technologies have been used for tamper-resistant storage of logging keys, performing cryptographic operations, and providing evidence of platform integrity to third-party verifiers using remote attestation.    

However, the advent of low-cost, mass-produced Internet of Things (IoT) devices complicates the use of trusted computing for tamper-resistant logging.   
Numerous proposals suggest using IoT devices for remote health monitoring~\cite{patel}, identifying fires and gas leakages~\cite{chen:home}, and detecting falls and injuries in the homes of the elderly and disabled~\cite{rashidi} -- all of which are natural applications for tamper-resistant logging.  Unfortunately, discrete hardware TPMs, which underpin many existing proposals, cannot directly host arbitrary applications without additional processes, such as launching and locally attesting applications from a TPM-backed virtual machine~\cite{bock,perez2006vtpm}.  Including such processes within the device's Trusted Computing Base (TCB) -- the set of software and hardware components essential to its security -- widens the scope for introducing security and performance defects~\cite{mccune,singareducing}.  One promising solution is the Trusted Execution Environment (TEE), which offers TPM-like functionality alongside strong isolated execution of critical applications, while using the core execution hardware of conventional operating systems.  TEEs have become widely-deployed in recent years, notably in the form of Intel Software Guard eXtensions (SGX) and TEEs built on ARM TrustZone.  Indeed, Trustonic estimated that one billion devices contained their TrustZone-based TEE alone in early 2017~\cite{trustonic}.  However, TEE-based logging schemes -- discussed in Section \ref{sec:related} -- have hitherto applied only server-side TEEs to protect logs transmitted from remote devices.

In this paper, we present EmLog, which leverages the GlobalPlatform TEE and ARM TrustZone for protecting logs \emph{at source} on mobile and embedded systems.  EmLog offers further security benefits over past work, including public verifiability of log origin, resilience to TEE key compromise, and supports secure I/O with peripheral devices.  After reviewing related work (Section \ref{sec:related}), we formalise the requirements and threat model in Section \ref{sec:reqs}.  EmLog is implemented on an off-the-shelf ARM development board hosting OP-TEE~\cite{optee} -- an open-source and GlobalPlatform-compliant TEE that uses TrustZone (Section \ref{sec:imp}) -- and evaluated using three datasets in Section \ref{sec:eval}.  Finally, we conclude our work in Section \ref{sec:conc} and identify future areas of research.  To our knowledge, this is the first attempt at preserving logs on constrained devices using a standardised TEE. The contributions of this paper are: \textbf{1)}, the development of a novel secure logging scheme for creating tamper-resistant logs with trust assurances, tailored for ARM-based constrained devices, like wearables and sensing platforms; and \textbf{2)}, a test-bed implementation using a GlobalPlatform-compliant TEE that uses ARM TrustZone, with performance benchmarks across three datasets.  The results indicate that EmLog has low run-time memory footprint, five-times persistent storage overhead, and 430--625 logs/sec throughput.
\section{Related Work}
\label{sec:related}
Existing proposals may be categorised as: \textbf{1)}, \emph{secure untrusted system logging}, focusing on cryptographic methods for detecting tampered logs on untrusted platforms; and \textbf{2)}, \emph{trusted logging}, for applying trusted hardware primitives for log preservation.  We briefly examine key proposals and their contributions.

\subsection{Secure Untrusted System Logging}
\label{sec:seclog}

Schneier and Kelsey~\cite{schneier} propose the use of MACs with linear one-way hash chains to protect log integrity.  Each chain entry is found by successively hashing the log content with the previous log's hash, which is accompanied by a MAC keyed under the hash of the previous MAC key.  The initial key is a pre-shared key (PSK) between the logging device and a trusted verifier, which allows the MAC hash chain to be recomputed and verified.  Bellare and Yee~\cite{bellareyee} propose a similar scheme using the formalised notion of \emph{forward integrity} in which it is computationally infeasible to alter past entries after a key compromise.  This is achieved by updating the secret key at regular time intervals (epochs) using an update process based on a chain of pseudo-random functions to key the log MACs in each epoch.  Holt~\cite{holt} proposed Logcrypt, which uses public-key cryptography alongside MACs to achieve \emph{public verifiability}, so third-parties can authenticate the origin of log entries without knowledge of a secret PSK -- shortfalls of \cite{schneier} and \cite{bellareyee}.  Ma et al.~\cite{ma:new} introduce FssAgg, which uses an aggregated chain of signatures to achieve public verifiability and to thwart \emph{truncation attacks}, where an attacker aims to delete a tail-end subset of log entries.  Yavuz et al.~\cite{yavuz} proposed LogFAS, which addresses both challenges with better storage and computational complexity than \cite{holt} and \cite{ma:new} using the Schnorr signature scheme.  Recently, Hartung~\cite{hartung2017attacks} presented four attacks against LogFAS~\cite{yavuz} and two variants of FssAgg~\cite{ma:new}, which enables secret key recovery and log forgery; as a result, both schemes are dissuaded from use.

\subsection{Secure Logging with Trusted Hardware}
\label{sec:trlog}

Early work by Chong et al.~\cite{chong:tpm} explored trusted hardware (Java iButton) to protect the initial PSK of the Schneier and Kelsey scheme~\cite{schneier}.  Later, Sinha et al.~\cite{sinha:tpm} suggested a using a TPM with a forward integrity scheme based on branched key chaining.  Logs are divided into epochs (blocks), each comprising a sequence of hash-chained log entries (sub-epochs).  The root entries of each epoch are hash-chained with past epochs, which creates a two-dimensional hash chain to prevent \emph{re-ordering attacks} in which an attacker re-orders log blocks to mislead auditors.  For each new epoch, the previous epoch's logs are securely stored using the TPM's seal functionality, which encrypts the logs with a TPM-bound key so only that particular TPM can decrypt/`unseal' them.  B{\"o}ck et al.~\cite{bock} explore the use of AMD's Secure Virtual Machine (SVM) -- an early inception of the TEE -- for launching a \texttt{syslog} client daemon and logging application from the TPM's secure boot chain.  The logger executes with access to TPM-bound key-pairs for encrypting and signing log entries.  Upon request, the logs are decrypted and transmitted to the verifying party; the TPM keys are certified for authenticating that signed logs originated from the SVM.

Nguyen et al.~\cite{cloudlogger} propose streaming medical logs to a server application in Intel SGX  (see Section \ref{sec:tees}) that applies the tamper-resistance.  Logs are sent to the Intel SGX application (`enclave') over TLS, which computes a hash chain comprising a signature of each record; TPMs are used to authenticate the medical devices to the server, and on the server's end to securely store log hash chains using its sealing mechanism.
Karande et al.~\cite{karande} introduce SGX-Log, which protects server-side device logs received from remote devices.  SGX-Log, like \cite{sinha:tpm}, uses block-based hash chains with SGX's secure storage for log integrity and confidentiality. The authors note that continual sealing also provides resilience to attacks in which large volumes of logs in memory are lost due to an unauthorised power loss.   Remote attestation is also suggested to authenticate the server enclave before transmitting the logs.  The proposed scheme is evaluated using three datasets, yielding a small ($<7\%$) overhead versus a non-SGX implementation. 

\subsection{Discussion}

Modern TPM- and TEE-based approaches~\cite{cloudlogger,karande,sinha:tpm,bock} still fall short of satisfying many desirable properties identified in past work.  Public verifiability of origin, as in \cite{bock}, has not been addressed in recent TEE loggers, which could be potentially useful to authenticate system data from remote devices, e.g. generating trust scores from log data for access control \cite{bao:iot} and continuous authentication \cite{micallef2015sensor,shepherd2017towards}. 
Recent TEE-based schemes, i.e.\ \cite{karande} and \cite{cloudlogger}, focus primarily on protecting logs \emph{after} being received by a server-side log processing application; an attacker on the source device may simply tamper the logs before reaching the server that applies some tamper-resistance algorithm.
To complicate matters, source devices are unlikely to transmit logs in real-time to minimise network and computational overhead, and so secure storage methods should be used to preserve unsent logs.  Additionally, TEEs typically contain other security-critical applications, e.g. for fingerprint matching (as in Android\footnote{\scriptsize{https://source.android.com/security/authentication/fingerprint-hal}}) and payment tokenisation (see Samsung Pay\footnote{\scriptsize{http://developer.samsung.com/tech-insights/pay/device-side-security}}).  As a result, a TEE-based logging mechanism should operate with reasonable resource consumption, e.g. run-time memory, to limit the rise of Denial of Service (DoS) conditions.   

\section{Trusted Execution Environments (TEEs)}
\label{sec:tees}
GlobalPlatform defines a TEE as an isolated execution environment that \emph{``protects from general software attacks, defines rigid safeguards as to the data and functions a program can access, and resists a set of defined threats''}~\cite{gp:tee}.  TEEs aim to isolate applications from integrity and confidentiality attacks from a conventional operating system.  Applications in the conventional OS and TEE -- referred to as `untrusted' and `trusted' worlds respectively in GlobalPlatform nomenclature -- reside in separate memory address spaces, and trusted hardware is used to monitor and prevent unauthorised memory accesses from the untrusted world.  TEE applications may allocate shared memory spaces or expose predefined functions via an API mediated by a high-privilege secure monitor.  Next, we summarise the leading commercial TEE architectures.

\begin{figure}
\centering
\includegraphics[width=0.75\linewidth]{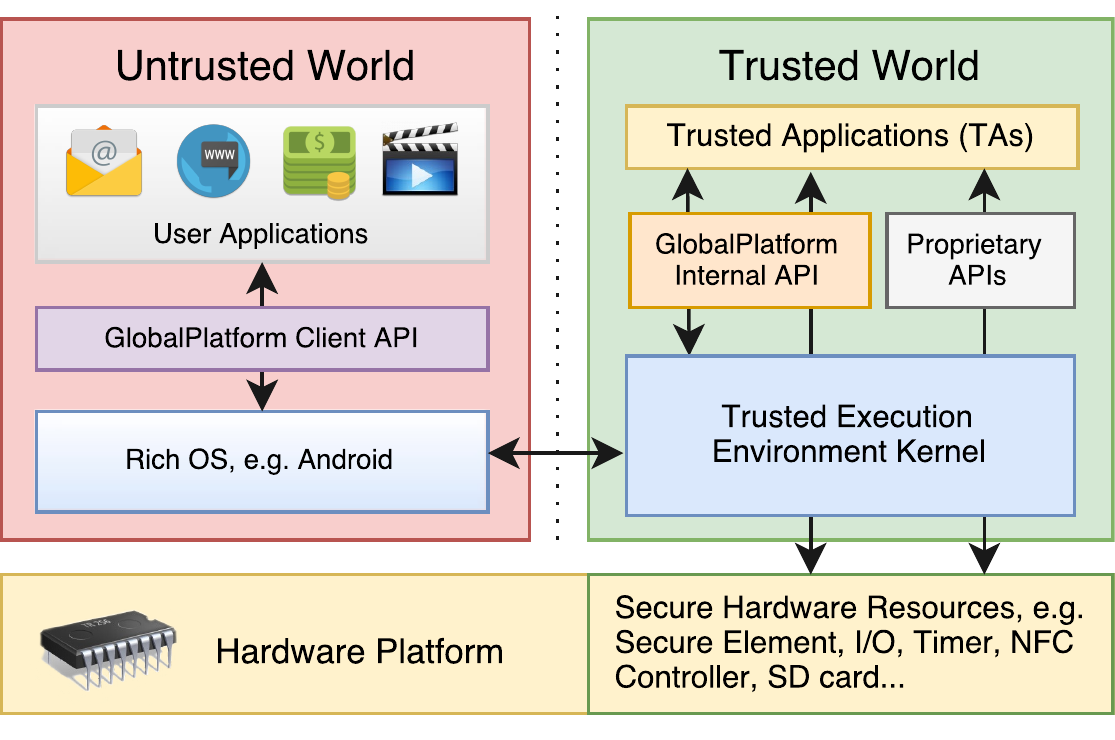}
\caption{GlobalPlatform TEE architecture.}
\label{fig:gptee}
\end{figure}

The \textbf{GlobalPlatform (GP) TEE} maintains two worlds for all trusted and untrusted applications. A TEE-based kernel is used for scheduling, memory management, cryptographic methods and other basic OS functions; TEE-resident Trusted Applications (TAs) may access OS functions exposed by the GP TEE Internal API (see Figure \ref{fig:gptee}). 
The GP TEE Client API \cite{gp:tee} defines the interfaces for communicating with TAs from untrusted world applications.  The GP specifications also cover the use of external secure elements (GP Secure Element API), secure storage, and networking (GP Sockets API) \cite{gp:arch}.   One method for instantiating the GP TEE is using ARM TrustZone, which enables two isolated worlds to co-exist in hardware.  This is achieved using two virtual cores for each world per physical CPU core and an extra CPU bit, the NS bit, for distinguishing between untrusted/secure world execution modes.  TrustZone provides secure I/O with peripheral devices connected over standard interfaces, e.g. SPI and GPIO, by routing interrupts to the TEE OS.  This is performed via the TrustZone Protection Controller (TZPC), responsible for securing on-chip peripherals, and the TrustZone Address Space Controller (TZASC) for protecting memory-mapped devices from untrusted world accesses.

\textbf{Intel Software Guard Extensions (SGX)} is an extension to the X86-64 instruction set that enables on-demand creation of `enclaves' per application.  Enclaves reside in isolated memory regions within RAM with accesses mediated by the CPU, which is considered trusted~\cite{sgx:costan}.  Enclaves may access the memory space of a regular OS, but not vice-versa, and enclaves cannot access other enclaves arbitrarily.  Like TPMs, SGX offers secure storage through `sealing' in which data is encrypted and made accessible only to that enclave.  Remote attestation enables third-party verification of enclaves and secret provisioning using Enhanced Privacy ID (EPID) -- a Direct Anonymous Attestation (DAA) protocol by Brickell et al.~\cite{brickell:epid}.   SGX has been supported from the release of the Skylake microarchitecture (from 2015).

Despite some high-level similarities, SGX is not GlobalPlatform-compliant.  Intel SGX is currently restricted solely to Intel CPUs, while the GP TEE is typically deployed on ARM System-on-Chips (SoCs) using TrustZone, as used by many IoT devices, e.g. Raspberry Pi 3\footnote{\scriptsize https://www.raspberrypi.org/products/}, NEST thermostat\footnote{\scriptsize https://nest.com}, and 95\% of consumer wearables according to ARM~\cite{arm:wearable}.  The reader is referred to \cite{shepherd:tee} for a detailed survey of secure and trusted execution environments for IoT devices.

\section{System Requirements}
\label{sec:reqs}

We formalise the requirements for a TEE-based system for protecting logs on constrained devices.  The proposal should satisfy the following security and functional requirements drawn from the issues identified in Section \ref{sec:related}:

\begin{enumerate}
\item[R1.] \emph{Isolated execution}: the system shall process logs in an environment isolated from a regular `rich' OS, e.g. Android, to provide strong integrity assurances of the application and data under execution.
\item[R2.] \emph{Forward integrity}: the integrity of a given block of logs shall not be affected by a key comprise of a previous block.
\item[R3.] \emph{Log confidentiality}: on-device log confidentiality should be preserved to prevent the disclosure of potentially sensitive entries.
\item[R4.] \emph{Remote attestation}: the proposal shall allow third-parties to verify the logging application's integrity post-deployment to provide assurances that logs were sourced from an integral and authentic platform.
\item[R5.] \emph{Secure log retrieval}: remote, authorised third-parties shall be able to securely retrieve device logs with mutual trust assurances.
\item[R6.] \emph{Public verifiability}: the system shall allow third-parties to authenticate the origin of log entries without access to private key information.
\item[R7.] \emph{Truncation attack-resistant}: the system shall be resistant to attacks that aim to delete a contiguous subset of tail-end log entries.
\item[R8.] \emph{Re-ordering attack-resistant}: the proposal shall resist attempts to change the order of entries in the log sequence.
\item[R9.] \emph{Power-loss resilience}: the loss of tamper-resistant logs shall be minimised in the event of a device power-loss.
\item[R10.] \emph{Suitable root of trust}: a root of trust for constrained device architectures shall be used, ideally without requiring additional security hardware.
\end{enumerate}

The threat model considers two adversary types:

\begin{itemize}
\item \emph{On-device software adversary}: a software-based attacker that compromises the system at time $t$ and attempts to arbitrarily alter, forge or observe logs produced before $t$.  This may operate at any protection level in the untrusted world, i.e.\ Rings 0--3, including arbitrarily altering execution flow and accessing non-TEE kernel space services. 
\item \emph{Network adversary}: an adversary that attempts to arbitrarily alter, forge, replay or observe logs between the source device and the verifier over a network channel, e.g. WiFi/802.11.  The attacker may also attempt to masquerade as a legitimate party to either end-point to collect logs illicitly.
\end{itemize}

Like past work, we trust the TEE and do not attempt to secure untrusted world logs \emph{after} a compromise after time $t$, since a kernel-mode adversary may simply read/write directly to the kernel message buffer used to queue log entries (see Section \ref{sec:proc}).   We also consider hardware and related side-channel attacks, e.g. power analysis, beyond the scope of this work, as these threats fall outside the security remit of TEEs.  The reader is referred to the GlobalPlatform TEE Protection Profile~\cite{gp:tee} for a specification of their protection scope.

\section{EmLog Architecture Design}

\begin{figure}
\centering
\includegraphics[interpolate=true,width=\linewidth]{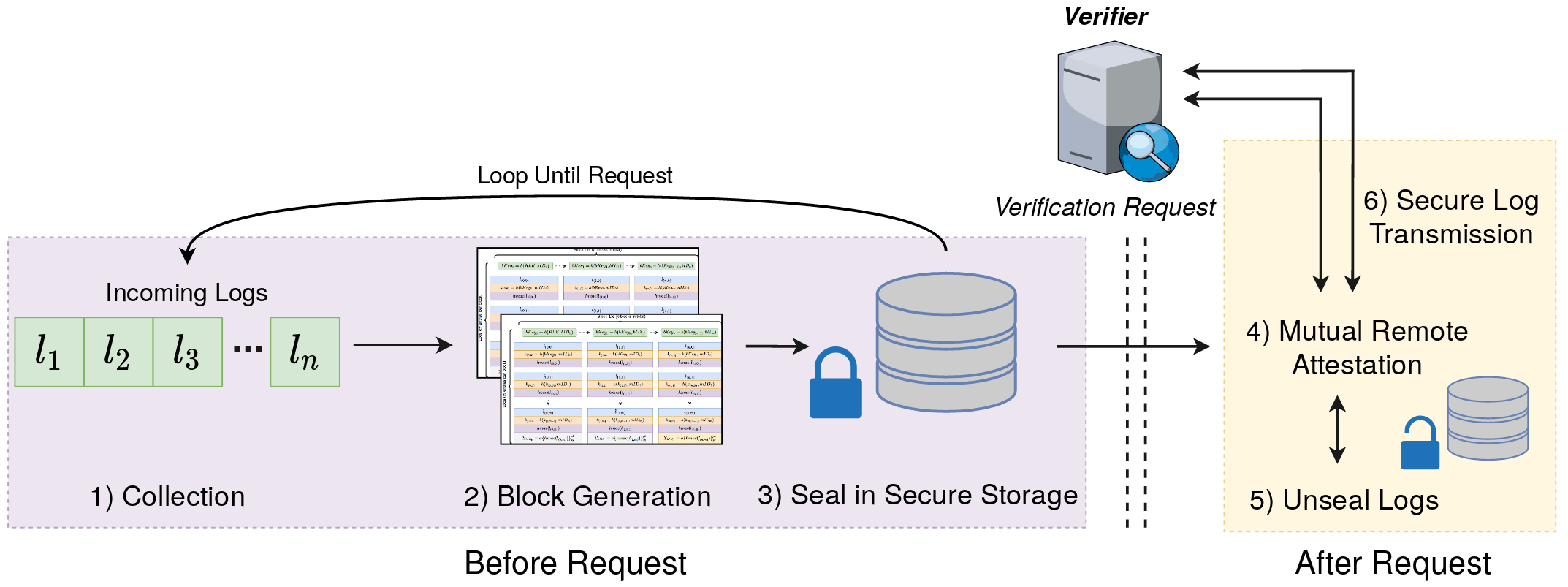}
\caption{High-level TEE-based logging workflow.}
\label{fig:flow}
\end{figure}

We assume the presence of a GlobalPlatform-compliant TEE, a service provider that provisions EmLog into the TEE before deployment, and a third-party wishing to retrieve all or a partial set of the device's logs.
The GP TEE, which maintains two sets of applications for each world, necessitates two logging components: one that collects logs from untrusted world applications and transmits these to the TEE over the GP Client API, and another that applies the protection algorithm within the TEE and responds to retrieval requests. 
An extension to the hash matrix in \cite{sinha:tpm} and \cite{karande} is proposed to apply the tamper-resistance scheme within the GP TEE, which achieves integrity protection and public verifiability (Section \ref{sec:block}).   
Next, the log blocks are stored every $n$ blocks, or at a time epoch $t$, using the secure storage functionality of the GP TEE.  After receiving a retrieval request, the source TEE authenticates the remote verifier and vice-versa, after which the blocks are unsealed and transmitted over a secure channel between the TEEs (Section \ref{sec:storage}).   We illustrate this process in Figure \ref{fig:flow}.

\subsection{Log Collection}
\label{sec:proc}

A conventional (Linux) kernel uses an internal message ring buffer to store log messages, which is made available to user space monitoring applications, such as \texttt{dmesg} and \texttt{klogd}, using the \texttt{sys\_syslog} syscall.  For user-mode logging, \texttt{syslogd} listens on \texttt{/dev/log}, where logs are registered to using the \texttt{syslog} function from the C standard library.  Logs are subsequently written to file or transmitted through the \texttt{syslog} protocol to a remote server over UDP.  Some implementations, e.g. \texttt{syslog-ng}, provide further functionality like streaming logs over TCP with TLS.  For collecting untrusted world logs, we suggest a \texttt{syslogd} variant that transmits logs to the EmLog TA within the GP TEE via the GP Client API.

\subsection{Block Generation}
\label{sec:block}

We propose a variant of the hash matrix used in \cite{karande} and \cite{sinha:tpm} for log sequence integrity.  Here, hash sequences are created in which each block key, $bK$, is derived using a one-way hash function, $h$, over the previous block key and current block ID, $bID$; that is, $bK_{bID} = h(bK_{bID-1}, bID)$.   The initial block key ($bID=0$) is derived from a device-specific Root Logging Key (RLK).  Each block key is used to derive an individual message key, $k$, for keying an HMAC in a similarly chained fashion, i.e. $k_{(bID,mID)} = h(k_{\{bID,(mID-1)\}}, mID)$ for log entry $mID$ in block $bID$, up to the block size $m$.  Note that $bK$ is used to derive $k$ when $mID=0$.  A block-based approach provides power-loss resilience and truncation resistance (developed further in Section \ref{sec:storage}) while allowing the retrieval of subsets, i.e. blocks $i$ to $j$, without transmitting all logs from the genesis block ($bID=0$) to the remote verifier.

As it stands, this scheme is vulnerable to forgery attacks if just a single block key is compromised: an adversary can apply $h$ on the leaked key with the next block ID to forge subsequent blocks and entries therein.    Storing RLK and deriving keys within trusted hardware, e.g. an secure element (SE) or TPM, is desirable, but this adds hardware complexity to already-constrained devices with respect to raw component and integration costs.  TEEs provide strong resilience to software attacks, but, unfortunately, are not invulnerable to developer-induced programming and API errors.  The impact of RLK and block key divulgence, however, can be limited using key derivation, as described below.

  \begin{figure}
\centering\includegraphics[interpolate=true,width=\columnwidth]{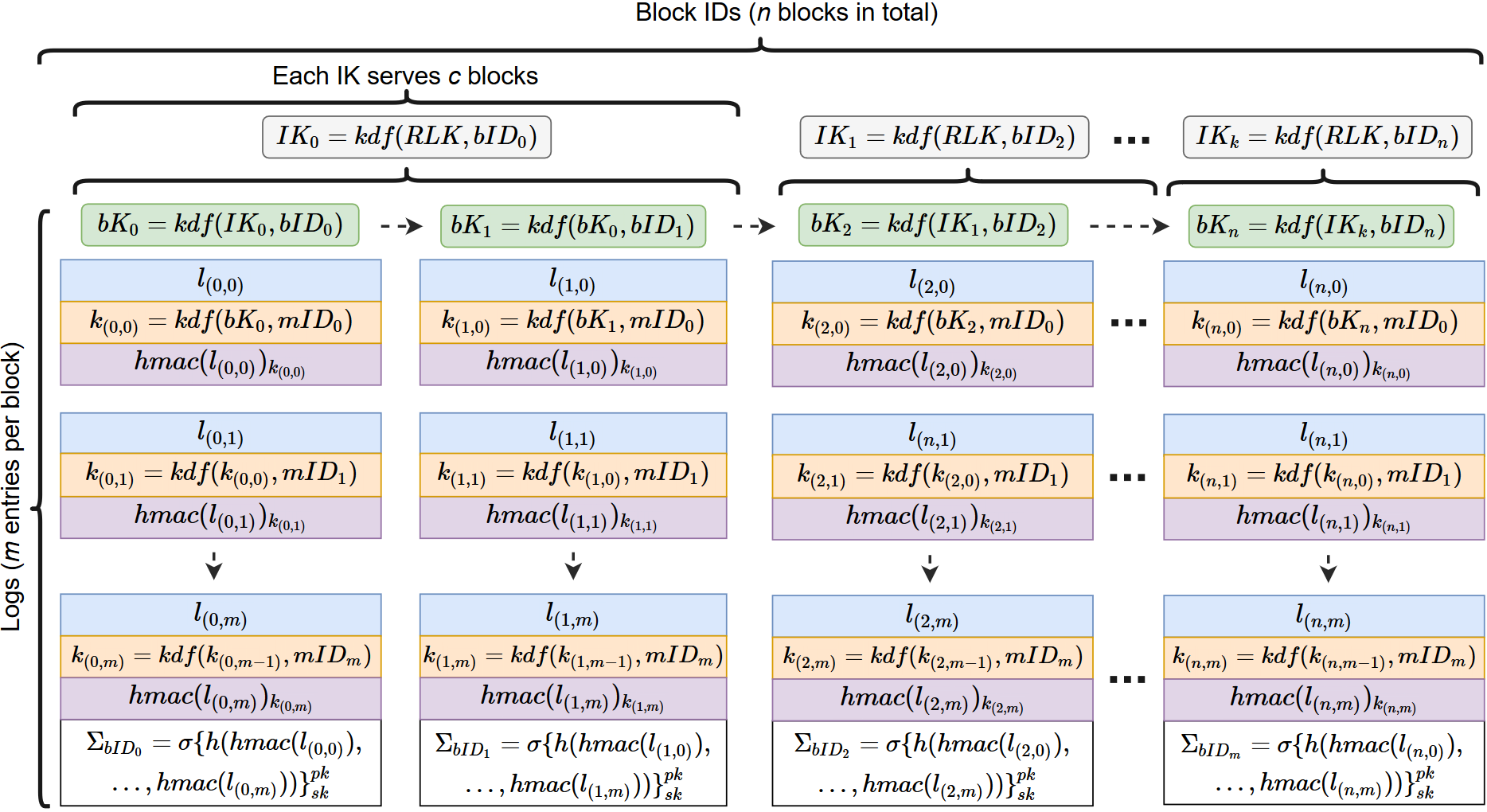}
\caption{Proposed two-dimensional, signature-based log structure.}
\label{fig:2d}
\vspace{-0.5cm}
\end{figure}

\textbf{Key Derivation.}
We suggest a simple scheme as follows: \textbf{1)}, intermediate keys (IKs) are derived from the RLK  using a secure key derivation function, each of which serves $c$ blocks; \textbf{2)}, each IK derives an initial block key, $bK_{bID}$, for that block group, before sealing the IK immediately to storage; \textbf{3)}, $bK_{bID}$ is used to generate the block's message-specific keys;  \textbf{4)}, the next $bK_{bID}$ is derived using $kdf(bK_{bID-1}, bID)$ for up to $c$ blocks, after which another IK is generated.  In past proposals, a block key disclosure would require re-provisioning RLK -- a device-specific, possibly hardware-infused key, which would affect the device in perpetuity without potentially costly intervention.  Our approach (Figure \ref{fig:2d}) limits the damage wrought by a compromised block key by affecting only future blocks \emph{in that group}.  In the worst case, besides divulging RLK, the exposure of IK can compromise only $c$ blocks at most.
  
For the key derivation function, we suggest the HMAC-based extract-and-expand KDF (HKDF) by Krawczyk~\cite{hkdf,hkdf:rfc} (RFC 5689).  HKDF takes keying material and a non-secret salt as input, and repeatedly generates HMACs under the input to return cryptographically strong output key material.  Unlike plain hash functions, used prevalently in past work, HKDF produces provably strong key material from as-strong or weaker input key material.

\textbf{Log Integrity and Verifiability.}
For log message integrity, first compute $hmac(\mathcal{\ell}_{(bID,mID)})$ for message $\mathcal{\ell}$ with block ID, $bID$, and message ID, $mID$, under key $k_{(bID, mID)}$ -- derived from the previous message key or, for $mID=0$, the block key.  Each $k$ should be immediately deleted from memory to limit memory consumption and exposure.  This does not prevent auditing log sequences, since message keys may be regenerated from the pre-shared RLK.    

In current symmetric-only schemes~\cite{chong:tpm,karande,schneier,sinha:tpm}, public verification of log origin (R6 in Section \ref{sec:reqs}) requires knowledge of block and message keys on all interested devices, derived ultimately from RLK.  Revealing RLK is evidently undesirable because it enables the malicious creation and manipulation of valid blocks.  Rather, we propose signing each block with an efficient signature scheme, $\sigma$, such as ECDSA, and a device-specific signing key-pair $(pk, sk)$ over the concatenation of the block message HMACs (Figure \ref{fig:2d}).  This key-pair should be certified to provide data origin authentication.  The RLK and key-pair should be accessible only to the TEE, which is achievable using the TEE's secure storage mechanism or, for hardware tamper-resistance (with its complexities), using an external SE as suggested by GlobalPlatform~\cite{gp:tee}.  In some circumstances, logs may contain sensitive data, in which case we suggest limiting verifiability to whitelisted entities, e.g. devices from the same manufacturer or service provider.  It is also observed that the block size, $m$, is inversely proportional to the number of signing operations; smaller block sizes will incur more signing operations for a given set of log entries (see Section \ref{sec:eval} for this overhead).

\subsection{Secure Storage and Remote Retrieval}
\label{sec:storage}

Real-time log streaming is likely to be detrimental for power- and network-limited devices, and we suggest storing blocks prior to eventual transmission within the TEE's secure storage.  Secure storage can be implemented in two ways according to GlobalPlatform: \textbf{1)}, using the file system and storage medium, e.g. flash drive, controlled by the untrusted world \emph{``as long as suitable cryptographic protection is applied, which must be as strong as that used to protect the TEE code and data itself''}~\cite{gp:internal}.  Or \textbf{2)}, using hardware controlled only by the TEE, e.g. an external SE.  Method \textbf{2)} is resilient against adversaries that aim to delete encrypted records from the file system\footnote{Note that, in general, arbitrary log deletion is difficult to prevent robustly without dedicated WORM (Write-Once, Read-Many) storage.}, but naturally requires additional security hardware. 
For method \textbf{1)}, log blocks are sealed using authenticated encryption (AES in GCM mode) with a key derived specifically for the TA under execution from a separate, device-specific root storage key.   This prevents other TAs or other entities from accessing secured data, thus providing on-device log confidentiality (R3), integrity and authenticity.   

Securely storing every completed block, i.e. in \cite{karande}, may yield undesirable performance overhead for the devices targeted in this work.  Rather, the parameters $c$ (block group size) and $m$ (block length) can control the number and size of blocks kept in RAM respectively. This satisfies truncation attack-resistance (R7) and power-loss resilience (R9), in addition to R3, by limiting the number of new blocks kept in memory (for sufficiently small values of $c$ and $m$).

In past work, log retrieval is proposed using TLS~\cite{cloudlogger}, or one-way remote attestation for authenticating the platform of the remote verifier~\cite{karande}.  (Many remote attestation protocols, e.g. \cite{brickell:epid}, typically enable secure channels to be bootstrapped, over which unsealed logs can be transmitted securely).  However, the remote authority, which may itself process logs in its own TEE \cite{cloudlogger,karande}, is likely to request reciprocal trust assurances from the source TEE, i.e.\ remote attestation for both the source \emph{and} verifying entities.  Rather than performing one-way attestation separately for both entities, one alternative is mutual TEE attestation \cite{shepherd:ares} in which both communicating TEEs are attested and authenticated within the protocol run.  Similarly, a secure channel can be bootstrapped from \cite{shepherd:ares} between the TEE end-points over which unsealed logs can be transmitted securely without exposing them to untrusted world elements. 

\section{Implementation}
\label{sec:imp}

We implemented EmLog using OP-TEE -- an open-souce, GlobalPlatform compliant TEE by Linaro~\cite{optee} -- with Debian (Linux) as the untrusted world OS.  An untrusted world application was developed for collecting log entries from file, which were sent subsequently to the EmLog TA using the GP Client API~\cite{gp:tee}.  Each entry was processed into a data structure comprising a 4-byte message ID, 32-byte HMAC (SHA-256) tag, and 256-byte field for the entry text.  The GP Internal API~\cite{gp:internal} was used to interface with the cryptographic and secure storage methods; in OP-TEE, cryptographic methods are implemented using the \texttt{LibTomCrypt} library, and we opted for secure storage in which data is encrypted to the untrusted world file system (residing on 32GB eMMC flash memory).   256-bit ECDSA (NIST \texttt{secp256r1} curve) was used to sign each block, which was placed into a separate data structure comprising the processed messages and a 4-byte block ID; currently, only the NIST curves are defined in the GP TEE specifications.    The GP Internal API defines its own memory allocation functions, i.e. \texttt{TEE\_Malloc} and \texttt{TEE\_Free}, for dynamically (de-)allocating memory to regions accessible only to the TA, which were used frequently for memory-managing blocks and messages at run-time.

Unsurprisingly, memory consumption quickly became problematic when working with large datasets (discussed in Section \ref{sec:eval}).  For the current OP-TEE release, 32MB RAM is allocated for the TEE kernel and all resident TAs, with the rest allocated to the untrusted world OS. For a standard TA, the Linaro Working Group\footnote{\scriptsize https://wiki.linaro.org/WorkingGroups/Security/OP-TEE} stipulates a default stack and heap size at 1kB (stack) and 32kB (data) respectively, both of which can be increased up to a maximum 1MB per TA. 

\section{Evaluation}
\label{sec:eval}

EmLog was evaluated using a HiKey LeMaker -- an ARM development board with a Huawei HiSilicon Kirin 620 SoC with 2GB RAM and an ARM Cortex A53 CPU (eight-cores at 1.2 GHz with TrustZone extensions).  Such specifications are typical of modern medium-to-high end IoT-type systems, such as the Raspberry Pi 3 and Nest Thermostat.  The proposal was benchmarked using three log file datasets, described briefly:
\begin{enumerate}
\item \emph{U.S. Securities and Exchange Commission (SEC) EDGAR}:  Apache logs from access statistics to SEC.gov.  We use the latest dataset\footnote{\scriptsize http://www.sec.gov/dera/data/Public-EDGAR-log-file-data/2016/Qtr2/log20160630.zip} with over a million entries (192 MB). (Mean entry length: 115.08 characters; S.D.: 5.73).
\item \emph{Mid-Atlantic Collegiate Cyber Defense Competition (CDC)}: IDS logs from the U.S. National CyberWatch MACCDC event, with $\sim$166,000 (27 MB) of Snort fast alert logs\footnote{\scriptsize http://www.secrepo.com/maccdc2012/maccdc2012\_fast\_alert.7z}.  (Mean entry length: 165.27 characters; S.D.: 38.21).
\item \textbf{\emph{EmLogs}}: Our dataset from OP-TEE OS boot, initialisation and GlobalPlatform test suite logs via the \texttt{xtest} command, and untrusted world logs from \texttt{dmesg}.  Over 25,000 records (1.7MB). (Mean entry length: 94.14; S.D.: 49.33).  
\end{enumerate}

\begin{figure}[ht]
\centering
\includegraphics[interpolate=true,width=\columnwidth]{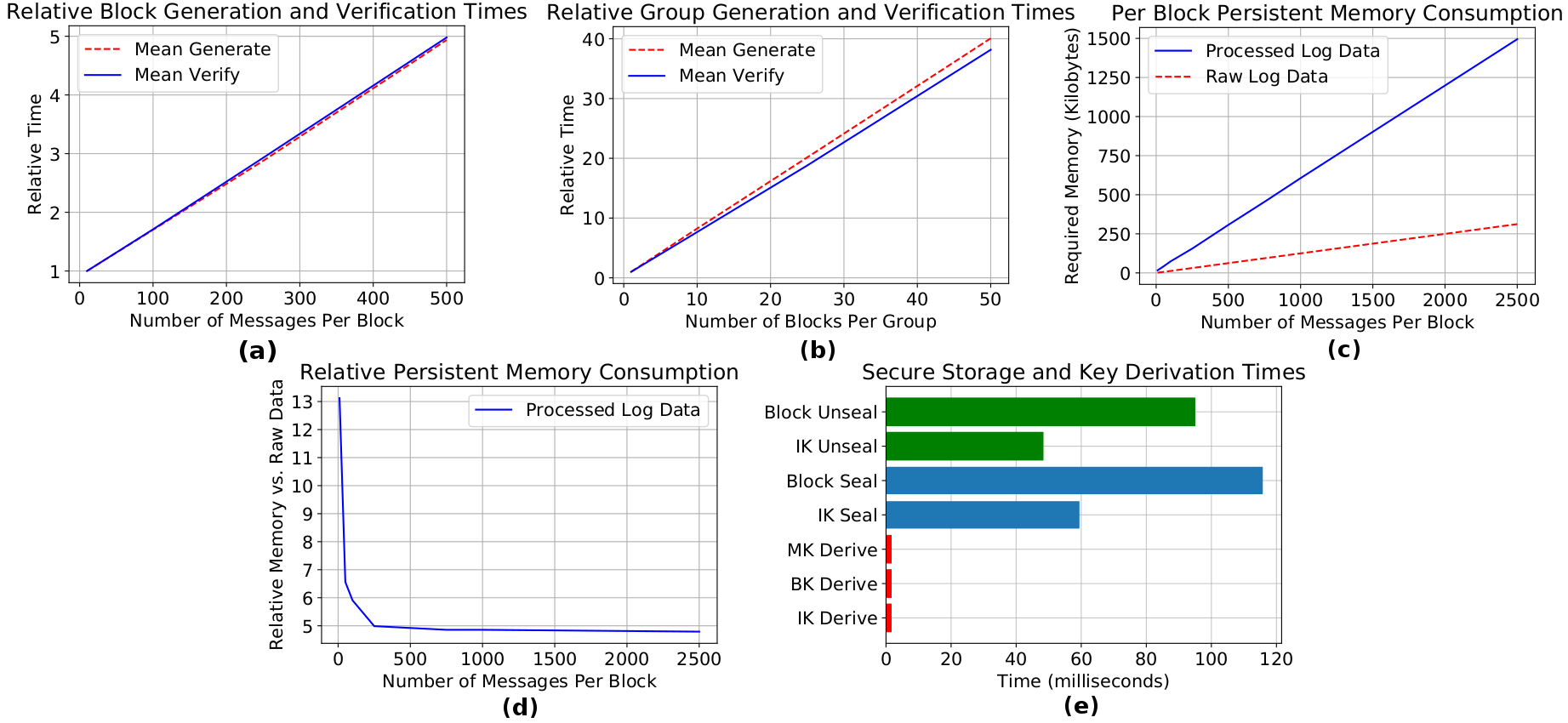}
\caption{\textbf{(a)}, Relative block creation and verification times versus block length; \textbf{(b)}, relative group generation and verification for varying numbers of blocks; \textbf{(c)}, persistent memory consumption for per block secure storage; \textbf{(d)}, relative memory consumption for group secure storage; and \textbf{(e)}, raw key derivation and secure storage times.}
\label{fig:perf}
\end{figure}

The results are shown in Tables \ref{tab:keyresults} to \ref{tab:memuse} and Figure \ref{fig:perf}.  Table \ref{tab:keyresults} shows the mean CPU time to derive 256-bit IKs, block and message keys from a pre-generated RLK using HKDF.  These were measured over 1,000 iterations within the EmLog TA using the GlobalPlatform \texttt{TEE\_Time} method for system time, implemented using the ARM Cortex \texttt{CNTFRQ} (CPU frequency) and \texttt{CNTPCT} (count) timing registers.  Table \ref{tab:keyresults} shows the mean time for sealing and unsealing IKs and blocks (for $m=100$, averaged across all entries) via the GP Internal API.  Table \ref{tab:hmacsig} lists the mean 256-bit ECDSA and HMAC-SHA256 times computed across all entries, while Table \ref{tab:blockresults} shows the mean creation and verification times of message blocks for each dataset (for varying values of $m$, the number of entries per block), as well as block groups.  In this context, verification encompasses the time to reconstruct the hash matrix in Figure \ref{fig:2d} and to verify the block signatures and message HMACs.  Group creation and verification time was measured for varying values of $c$ (blocks per group), which included the time for sealing and unsealing blocks to secure storage respectively.  Table \ref{tab:memuse} shows the mean persistent memory consumption of logs in secure storage, which was measured directly from \texttt{/data/tee} in the untrusted world file system, where OP-TEE stores sealed TA files.  Lastly, Figure \ref{fig:perf} shows the relative performance of secure storage, key derivation and block and group creation/verification times from Table \ref{tab:keyresults}.

\begin{table}
\centering
\caption{Mean key derivation and secure storage times (milliseconds; S.D. in brackets).}
\label{tab:keyresults}
\resizebox{\linewidth}{!}{
\begin{tabular}{|c|c|c|c|c|c|c|}
\hline
\multicolumn{3}{|c|}{\textbf{Key Derivation}}                                   & \multicolumn{2}{c|}{\textbf{Secure Storage Seal}} & \multicolumn{2}{c|}{\textbf{Secure Storage Unseal}} \\ \hline
\textbf{IK} & \textbf{Block Key} & \textbf{Message Key} & \textbf{IK}      & \textbf{Block} &  \textbf{IK}   & \textbf{Block}    \\ \hline
1.530 (0.067)  &  1.541 (0.062)  &  1.547 (0.088)  &  59.46 (3.78)  & 115.8 (5.36) & 48.22 (2.73) & 94.88 (2.80) \\ \hline
\end{tabular}
}
\end{table}

\vspace{-0.7cm}

\begin{table}
\centering
\caption{Mean HMAC and ECDSA generation and verification times  (milliseconds).}
\label{tab:hmacsig}
\resizebox{0.65\linewidth}{!}{
\begin{tabular}{|c|c|c|c|}
\hline
                & \textbf{HMAC (SHA-256)} & \multicolumn{2}{c|}{\textbf{ECDSA (NIST P256)}} \\ \hline
\textbf{Create} & 0.056 (0.020) & \multicolumn{2}{c|}{20.14 (1.29)}   \\ \hline
\textbf{Verify} & 0.059 (0.014)   & \multicolumn{2}{c|}{20.77 (1.33)}               \\ \hline
\end{tabular}
}
\vspace{-0.7cm}
\end{table}

\begin{table}
\centering
\caption{Mean block and group generation and verification times (milliseconds).}
\label{tab:blockresults}
\resizebox{\linewidth}{!}{
\begin{tabular}{|c|c|c|c|c|c|c|c|c|c|}
\hline
\multicolumn{2}{|c|}{\multirow{2}{*}{}}   & \multicolumn{4}{c|}{\textbf{Block}}                     & \multicolumn{4}{c|}{\textbf{Group } (fixed at $m=100$)}                    \\ \cline{3-10} 
\multicolumn{2}{|c|}{\textbf{Dataset}}                    & \textbf{($m=$)10} & \textbf{100} & \textbf{250} & \textbf{500} & \textbf{($c=$)1} & \textbf{10} & \textbf{25} & \textbf{50*} \\ \hline
\multirow{2}{*}{(1)}  & \textbf{Create} &   40.14 (2.1) & 70.45 (2.2) &  115.6 (2.6)  &  198.7 (3.6)  &    229.4 (6.4)   &   1898 (30.2)  &   4622 (48.1)  & 9101 (80.2)      \\ \cline{2-10} 
 & \textbf{Verify} &  41.18 (1.5) &  71.88 (1.7) & 114.8 (1.9) &  204.3 (2.1) &   213.4  (5.7)  &   1634 (23.7)   &   3967 (50.0) &  8302 (78.3)  \\ \hline
\multirow{2}{*}{(2)}  & \textbf{Create} & 39.84 (1.9) & 68.34 (1.7)  & 117.2 (1.7)  & 202.3 (2.0) &  231.7 (7.3)  &  1910	 (28.1)  &  4687 (64.0) &  9323 (81.5)  \\ \cline{2-10} 
                        & \textbf{Verify} & 40.05 (1.6) & 66.15 (1.8) & 119.9 (1.8) & 199.0  (2.0) &  215.2 (6.1) & 1658 (23.8) &  4046 (47.3) &      8165   (73.9)     \\ \hline
\multirow{2}{*}{(3)} & \textbf{Create} & 42.14 (3.0)  &  69.07 (2.1)   &   118.6 (2.3) & 201.9 (3.2) &  230.0 (6.9)    &  1890 (27.0)    &   4621 (42.8)           &  9274 (79.0) \\ \cline{2-10} 
                        & \textbf{Verify} &  40.01 (1.6) &  69.28 (1.8) & 120.4 (1.8) & 200.4  (1.9) & 217.3 (6.4) & 1656 (25.2) & 4132 (48.1) &  8188 (75.5)  \\ \hline
\end{tabular}
}
\begin{tablenotes}
\item * Heap size set to 2MB to accommodate all data.  All other experiments recorded with the maximum recommendation of 1MB.
\end{tablenotes}
\vspace{-0.4cm}
\end{table}

\begin{table}
\centering
\caption{Persistent memory consumption for per block secure storage (kilobytes).}
\label{tab:memuse}
\resizebox{0.7\linewidth}{!}{
\begin{tabular}{|c|c|c|c|c|c|c|c|}
\hline
\multicolumn{8}{|c|}{\textbf{Block Sizes}}                                                           \\ \hline
\textbf{($m=$)10} & \textbf{50} & \textbf{100} & \textbf{250} & \textbf{500} & \textbf{750} & \textbf{1000} & \textbf{2500} \\ \hline
16.38 & 40.96 &  73.73    &   155.65   &  307.20    &  454.66   &  606.21 & 1495.04 \\ \hline
\end{tabular}
}
\end{table}

\subsection{Discussion}

Little to our surprise, block generation and verification time scales linearly with message length, which, for large values of $m$, is influenced heavily by the key derivation operations (approximately $1.5$ms per message, shown in Table \ref{tab:keyresults}).  At smaller values, e.g. $m=10$, this is dominated mostly by the ECDSA overhead ($\sim$20ms, as per Table \ref{tab:hmacsig}).  Figure \ref{fig:perf} indicates that the relative timing overhead is $\sim$80--100\% for every 100 message increase in the block length.   

Group creation and verification times rise significantly with the number of blocks, $c$, kept in RAM before secure storage.  This is driven significantly by the secure storage overhead, which is measured at approximately $115.8$ms and $94.88$ms for sealing and unsealing respectively (Table \ref{tab:keyresults}).  Despite this, however, even the largest group sizes, $c=25$ and $c=50$ (2,500 and 5,000 entries in total), completed between 4.0 to 9.3 seconds, corresponding to a throughput of approximately 538 and 625 logs per second.  At first, it seems attractive to maximise $c$ to avoid the expense of secure storage operations, which caused the throughput to drop to $\sim$430 and 525 entries for the smallest groups ($c=1$ and $c=10$ blocks).  Maintaining many blocks in RAM, however, increases the impact of a power-loss; systems that log infrequently may see significant data loss if large numbers of logs spread over a large period of time are lost.  Consequently, $c$ should be set based on the expected log and transmission frequencies.

For memory consumption, all experiments were conducted within the Linaro Working Group's run-time recommendations (1MB stack and heap), except for $c=50$ blocks (5,000 entries), which required 2MB of each.  Expectedly, persistent memory consumption of block secure storage (Figure \ref{fig:perf}) scales linearly with message size.  Our test-bed uses a fixed 256-byte text field for each log entry, which accounts for the broadly similar performance across all datasets.  We also calculated the persistent memory consumption compared with the mean size of raw logs; the relative consumption is large for small block sizes ($m<250$), due likely to the fixed-size meta-data used by OP-TEE to manage cross-TA secure storage objects.  For larger block sizes, this converges to slightly under five-times overhead versus raw logs; the absolute size of smaller block sizes remains low, however, at 16--155 kilobytes, according to Table \ref{tab:memuse}.

\subsection{Requirements Comparison}

We compare the features of EmLog's with previous work in Table \ref{tab:comparison} using the requirements in Section \ref{sec:reqs}.  Notably, the use of ARM TrustZone and the GlobalPlatform TEE makes it appropriate for mobile and embedded devices targeted in this work (R10), unlike SGX-based schemes, which are restricted to Intel CPUs associated with laptop, desktop and server machines.  EmLog satisfies the features of related cryptographic and trust-based proposals, such as resistance to truncation (R7) and re-ordering (R8) attacks, and public verifiability of log origin (R6).  We also offer forward integrity protection for compromised block keys (R2) using more sophisticated key derivation, thus moving the cost-reward ratio further away from an attacker.  By avoiding TPMs, however, we relinquish strong hardware tamper-resistance, and we urge caution of our work in high-security domains, e.g. military and governmental use, where complex hardware and side-channel attacks are reasonable threats.  

\begin{table*}
\centering
\caption{Security requirements comparison of related work.}
\label{tab:comparison}
\resizebox{\linewidth}{!}{
\begin{threeparttable}
\begin{tabular}{@{}lccccccccccl@{}}
& R1 & R2 & R3 & R4 & R5 & R6 & R7 & R8 & R9 & R10 & Root of Trust \\ \midrule
 \textbf{Untrusted World Schemes} &  &  &  &  & & & & & & & \\
\emph{Schneier and Kelsey}~\cite{schneier} & -- & \xmark & \cmark & -- & -- & \xmark & \xmark & \xmark & -- & -- & -- \\
\emph{Bellare and Yee}~\cite{bellareyee} & -- & \cmark  & \cmark & -- & -- & \xmark & \xmark & \xmark & -- & -- & --\\
\emph{FssAgg}~\cite{ma:new} & -- & \cmark & \xmark & -- & -- & \cmark & \cmark & \cmark & -- & -- & -- \\
\emph{Logcrypt}~\cite{holt} & -- & \cmark & \cmark & -- & -- & \cmark & \xmark & \xmark & -- & -- & -- \\
\emph{LogFAS}~\cite{yavuz} & -- & \cmark & \xmark & -- & -- & \cmark & \cmark & \cmark & -- & -- & -- \\\midrule
\textbf{Trusted Logging} &  & &  & &  & & & & & & \\
\emph{Chong et al.}~\cite{chong:tpm} & \cmark & \xmark & \cmark & \xmark & \xmark & \xmark & \xmark & \xmark & \xmark & \xmark & Java iButton \\
\emph{Sinha et al.}~\cite{sinha:tpm} & \xmark &  \P & \cmark & \cmark & \xmark & \xmark & \cmark & \cmark  & \cmark & \xmark & TPM \\
\emph{B\"{o}ck et al.}~\cite{bock} & \cmark & \xmark & \P & \cmark & \xmark & \cmark & \xmark & \xmark & \xmark & \xmark & TPM \& AMD SVM \\
\emph{Nguyen et al.}~\cite{cloudlogger} & \cmark  & \xmark & \P & \cmark & \P & \xmark & \cmark  & \cmark  & \cmark & \xmark & Intel SGX \\
\emph{SGX-Log}~\cite{karande} & \cmark & \P & \cmark & \cmark & \P & \xmark & \cmark  & \cmark & \cmark & \xmark & Intel SGX \\\midrule
\textbf{EmLog} & \cmark & \cmark & \cmark & \cmark & \cmark & \cmark & \cmark  & \cmark & \cmark & \cmark & GlobalPlatform TEE \\\midrule
\end{tabular}
\begin{tablenotes}
\item \cmark -- Satisfies requirement; \xmark -- Does not satisfy; \P -- Partially satisfies; (--) -- N/A.
\end{tablenotes}
\end{threeparttable}
}
\end{table*}

\section{Conclusion}
\label{sec:conc}

In this paper, we introduced EmLog -- a tamper-resistant logging scheme for modern constrained device using the GlobalPlatform TEE and ARM TrustZone.  We began with a two-part review of related work in Section \ref{sec:related} by summarising cryptographic proposals and those reliant upon trusted hardware for tamper-resistant logging.  Next, the features of TEEs were assess in further detail in Section \ref{sec:tees}, before formulating the requirements and the threat model in Section \ref{sec:reqs} using past work.  After this, we introduced the architectural design and proposed an improved log preservation algorithm for providing public verifiability of log origin and key exposure resilience.  We described the implementation of EmLog in Section \ref{sec:imp} and presented indicative performance results using diverse datasets in Section \ref{sec:eval}.  For the first time, our work brings secure, TEE-based logging to mobile and embedded devices, and protects against strong software-based untrusted world and network adversaries.  Our evaluation shows that EmLog yields five-times persistent storage overhead versus raw logs for applying tamper-resistance; runs within reasonable run-time memory constraints for TEE applications, as stipulated by the Linaro Working Group; and has a throughput of up to 625 logs/sec.  In future work, we aim to investigate the following avenues:

\begin{itemize}
\item \emph{Group logging schemes for multiple devices}.  Expand EmLog to allow secure and efficient sharing of logs with nearby devices.  This could be used in schemes that compute trust scores prior to making group decisions~\cite{bao:iot}, e.g. authenticating users via contextual data from multiple wearable devices.
\item \emph{Privacy-preserving log usage}.  At present, devices that wish to use logs will receive raw logs, which may reveal privacy-sensitive data.  In future work, we aim to explore privacy-preserving methods for using logs without exposing raw entries to other devices.
\item \emph{TEE performance comparison}.  We hope to evaluate EmLog under other TEE instantiations, namely Intel SGX and other GP-compliant TEEs, such as TrustTonic's Kinibi, especially for micro-controllers on low-end IoT devices.
\end{itemize}

\section*{Acknowledgements}
Carlton Shepherd is supported by the EPSRC and the British government as part of the Centre for Doctoral Training in Cyber Security at Royal Holloway, University of London (EP/K035584/1).
The authors would also like to thank the anonymous reviewers for their valuable comments and suggestions.

\bibliographystyle{abbrv}
\bibliography{bib}

\begin{thebibliography}{10}

\bibitem{arm:wearable}
{ARM}.
\newblock {Markets: Wearables}, 2017.
\newblock https://www.arm.com/markets/wearables.

\bibitem{bao:iot}
F.~Bao and I.-R. Chen.
\newblock {Dynamic Trust Management for Internet of Things Applications}.
\newblock In {\em {International Workshop on Self-aware Internet of Things}},
  pages 1--6. ACM, 2012.

\bibitem{bellareyee}
M.~Bellare and B.~Yee.
\newblock {Forward Integrity for Secure Audit Logs}.
\newblock Technical report, Computer Science and Engineering Department,
  University of California at San Diego, 1997.

\bibitem{bock}
B.~B\"{o}ck, D.~Huemer, and A.~M. Tjoa.
\newblock {Towards More Trustable Log Files for Digital Forensics by Means of
  Trusted Computing}.
\newblock In {\em 24th International Conference on Advanced Information
  Networking and Applications}, pages 1020--1027. IEEE, 2010.

\bibitem{brickell:epid}
E.~Brickell and J.~Li.
\newblock {Enhanced Privacy ID from Bilinear Pairing for Hardware
  Authentication and Attestation}.
\newblock {\em International Journal of Information Privacy, Security and
  Integrity}, 1(1):3--33, 2011.

\bibitem{chen:home}
D.~Chen and M.~Wang.
\newblock {A Home Security ZigBee Network for Remote Monitoring Applications}.
\newblock In {\em International Conference on Wireless, Mobile and Multimedia
  Networks}, pages 1--4. IET, 2006.

\bibitem{chong:tpm}
C.~N. Chong, Z.~Peng, and P.~H. Hartel.
\newblock {Secure Audit Logging with Tamper-Resistant Hardware}.
\newblock In {\em Security and Privacy in the Age of Uncertainty: IFIP TC11
  18th International Conference on Information Security}, pages 73--84.
  Springer, 2003.

\bibitem{sgx:costan}
V.~Costan and S.~Devadas.
\newblock {Intel SGX Explained}.
\newblock {\em IACR Cryptology ePrint Archive}, 2016:86, 2016.
\newblock https://eprint.iacr.org/2016/086.pdf.

\bibitem{gp:tee}
{GlobalPlatform}.
\newblock {TEE Protection Profile (v1.2)}, 2014.

\bibitem{gp:internal}
{GlobalPlatform}.
\newblock {TEE Internal Core API (v1.1.1)}, 2016.

\bibitem{gp:arch}
{GlobalPlatform}.
\newblock {TEE System Architecture (v1.1)}, 2017.

\bibitem{hartung2017attacks}
G.~Hartung.
\newblock {Attacks on Secure Logging Schemes}.
\newblock {\em IACR Cryptology ePrint Archive}, 2017:95, 2017.
\newblock https://eprint.iacr.org/2017/095.pdf.

\bibitem{holt}
J.~E. Holt.
\newblock {Logcrypt: Forward Security and Public Verification for Secure Audit
  Logs}.
\newblock In {\em Proceedings of the 2006 Australasian Workshops on Grid
  Computing and E-research}, pages 203--211. Australian Computer Society, Inc.,
  2006.

\bibitem{iso}
{International Standards Organisation}.
\newblock {ISO/IEC 27001:20133 -- Information Technology, Security Techniques,
  Information Security Management Systems, Requirements}, 2013.
\newblock https://www.iso.org/standard/54534.html.

\bibitem{karande}
V.~Karande, E.~Bauman, Z.~Lin, and L.~Khan.
\newblock {SGX-Log: Securing System Logs With SGX}.
\newblock In {\em Proceedings of the 2017 Asia Conference on Computer and
  Communications Security}, ASIA CCS '17, pages 19--30, NY, USA, 2017. ACM.

\bibitem{nist:logs}
K.~Kent and M.~Souppaya.
\newblock {Guide to Computer Security Log Management}.
\newblock {\em NIST Special Publication}, 92, 2006.

\bibitem{hkdf}
H.~Krawczyk.
\newblock {Cryptographic Extraction and Key Derivation: the HKDF Scheme}.
\newblock In {\em Advances in Cryptology, 30th Annual Cryptology Conference
  (CRYPTO 2010)}, pages 631--648. Springer Berlin Heidelberg, 2010.

\bibitem{hkdf:rfc}
H.~Krawczyk and P.~Eronen.
\newblock {RFC 5869 -- HMAC-based Extract-and-expand Key Derivation Function
  (HKDF)}, May 2010.
\newblock https://tools.ietf.org/html/rfc5869.

\bibitem{optee}
Linaro.
\newblock {OP-TEE: Open Portable Trusted Execution Environment}, 2017.
\newblock https://www.op-tee.org/.

\bibitem{ma:new}
D.~Ma and G.~Tsudik.
\newblock {A New Approach to Secure Logging}.
\newblock {\em ACM Transactions on Storage}, 5(1):2, 2009.

\bibitem{mccune}
J.~M. McCune, Y.~Li, N.~Qu, Z.~Zhou, A.~Datta, V.~Gligor, and A.~Perrig.
\newblock {TrustVisor: Efficient TCB Reduction and Attestation}.
\newblock In {\em 2010 IEEE Symposium on Security and Privacy}, pages 143--158.
  IEEE, 2010.

\bibitem{micallef2015sensor}
N.~Micallef, H.~G. Kayac{\i}k, M.~Just, L.~Baillie, and D.~Aspinall.
\newblock {Sensor Use and Usefulness: Trade-offs for Data-driven Authentication
  on Mobile Devices}.
\newblock In {\em IEEE International Conference on Pervasive Computing and
  Communications}, pages 189--197. IEEE, 2015.

\bibitem{cloudlogger}
H.~Nguyen, B.~Acharya, R.~Ivanov, A.~Haeberlen, L.~T.~X. Phan, O.~Sokolsky,
  J.~Walker, J.~Weimer, W.~Hanson, and I.~Lee.
\newblock {Cloud-Based Secure Logger for Medical Devices}.
\newblock In {\em {IEEE 1st International Conference on Connected Health:
  Applications, Systems and Engineering Technologies}}, pages 89--94, June
  2016.

\bibitem{patel}
S.~Patel, H.~Park, P.~Bonato, L.~Chan, and M.~Rodgers.
\newblock {A Review of Wearable Sensors and Systems with Applications in
  Rehabilitation}.
\newblock {\em Journal of Neuro-engineering and Rehabilitation}, 9(1):21, 2012.

\bibitem{perez2006vtpm}
R.~Perez, R.~Sailer, L.~van Doorn, et~al.
\newblock {vTPM: Virtualizing the Trusted Platform Module}.
\newblock In {\em Proceedings of the 15th USENIX Security Symposium}, pages
  305--320, 2006.

\bibitem{rashidi}
P.~Rashidi and A.~Mihailidis.
\newblock {A Survey on Ambient-Assisted Living Tools for Older Adults}.
\newblock {\em IEEE Journal of Biomedical and Health Informatics},
  17(3):579--590, 2013.

\bibitem{schneier}
B.~Schneier and J.~Kelsey.
\newblock Secure audit logs to support computer forensics.
\newblock {\em ACM Transactions on Information and System Security (TISSEC)},
  2(2):159--176, 1999.

\bibitem{shepherd:ares}
C.~Shepherd, R.~N. Akram, and K.~Markantonakis.
\newblock {Establishing Mutually Trusted Channels for Remote Sensing Devices
  with Trusted Execution Environments}.
\newblock In {\em 12th International Conference on Availability, Reliability
  and Security (ARES)}. ACM, 2017.

\bibitem{shepherd2017towards}
C.~Shepherd, R.~N. Akram, and K.~Markantonakis.
\newblock {Towards Trusted Execution of Multi-modal Continuous Authentication
  Schemes}.
\newblock In {\em Proceedings of the 32nd Symposium on Applied Computing},
  pages 1444--1451. ACM, 2017.

\bibitem{shepherd:tee}
C.~Shepherd, G.~Arfaoui, I.~Gurulian, R.~P. Lee, K.~Markantonakis, R.~N. Akram,
  D.~Saveron, and E.~Conchon.
\newblock {Secure and Trusted Execution: Past, Present, and Future -- A
  Critical Review in the Context of the Internet of Things and Cyber-Physical
  Systems}.
\newblock In {\em 15th IEEE International Conference on Trust, Security and
  Privacy in Computing and Communications}, pages 168--177, 2016.

\bibitem{singareducing}
L.~Singaravelu, C.~Pu, H.~H{\"a}rtig, and C.~Helmuth.
\newblock {Reducing TCB Complexity for Security-Sensitive Applications: Three
  Case Studies}.
\newblock In {\em ACM SIGOPS Operating Systems Review}, volume~40, pages
  161--174. ACM, 2006.

\bibitem{sinha:tpm}
A.~Sinha, L.~Jia, P.~England, and J.~R. Lorch.
\newblock {Continuous Tamper-proof Logging Using TPM 2.0}.
\newblock In {\em 7th International Conference on Trust and Trustworthy
  Computing}, pages 19--36, NY, USA, 2014. Springer-Verlag.

\bibitem{trustonic}
Trustonic.
\newblock {Adoption of Trustonic Security Platforms Passes 1 Billion Device
  Milestone}, February 2017.
\newblock
  https://www.trustonic.com/news/company/adoption-trustonic-security-platforms-passes-1-billion-device-milestone/.

\bibitem{yavuz}
A.~A. Yavuz, P.~Ning, and M.~K. Reiter.
\newblock {Efficient, Compromise-Resilient and Append-Only Cryptographic
  Schemes for Secure Audit Logging}.
\newblock In {\em 2012 International Conference on Financial Cryptography and
  Data Security}, pages 148--163. Springer, 2012.

\end{thebibliography}

\end{document}